\documentclass{aa}
\usepackage{graphicx}
\usepackage{color}

\def\gtrsim{\mathrel{\hbox{\rlap{\hbox{\lower4pt\hbox{$\sim$}}}\hbox{$>$}}}}
\def\ltsim{\mathrel{\hbox{\rlap{\hbox{\lower4pt\hbox{$\sim$}}}\hbox{$<$}}}}

\begin{document}

\title{ 
Discovery of a weak magnetic field in the photosphere of the single giant Pollux
\thanks{Based on observations  obtained at the Canada-France-Hawaii Telescope (CFHT) which is operated by the National Research Council of Canada, CNRS/INSU and the University of Hawaii, and the T\'elescope Bernard Lyot (TBL) at Observatoire du Pic du Midi, CNRS/INSU and Universit\'e de Toulouse, France.}}

\author{M. Auri\`ere\inst{1}, G.A. Wade\inst{2}, R. Konstantinova-Antova\inst{3}, C. Charbonnel\inst{5,1}, C. Catala\inst{4},  W. W. Weiss\inst{6}, T. Roudier\inst{1}, P. Petit\inst{1},  J.-F. Donati\inst{1}, E. Alecian\inst{4}, R. Cabanac\inst{1}, S. Van Eck \inst{7}, C.P. Folsom\inst{8}, J. Power\inst{2}}
\offprints{M. Auri\`ere, {\tt michel.auriere@ast.obs-mip.fr}}

\institute{Laboratoire d'Astrophysique de Toulouse- Tarbes, Universit\'e de Toulouse, CNRS, 57 Avenue d'Azereix, 65000 Tarbes, France
\and
Department of Physics, Royal Military College of Canada,
  PO Box 17000, Station 'Forces', Kingston, Ontario, Canada K7K 4B4
\and
Institute of Astronomy, Bulgarian Academy of Sciences, 72 Tsarigradsko shose, 1784 Sofia, Bulgaria
\and
LESIA-UMR8109, CNRS and Universit\'e Paris VII, 5 place Janssen, F-92195 Meudon Cedex, France
\and
Geneva Observatory, University of Geneva, 51 Chemin des Maillettes, 1290 Versoix, Switzerland
 \and 
Institut f\"ur Astronomie, Universit\"at Wien, T\"urkenschanzstrasse 17, A-1180 Wien, Austria
\and
Institut d'Astrophysique, Universit\'e Libre de Bruxelles, Campus Plaine - C.P. 226, 1050 Bruxelles, Belgium
\and
Armagh Observatory, College Hill, Armagh, BT619DG, Northern Ireland, UK}

 \date{Received ??; accepted ??}

\abstract
 {}
{We observe the nearby, weakly-active single giant, Pollux, in order to directly study and infer the nature of its magnetic field.}
{We used the new generation spectropolarimeters ESPaDOnS and NARVAL to observe and detect circular polarization within the photospheric absorption lines of Pollux. Our observations span 18 months from 2007-2009. We treated the spectropolarimetric data using the Least-Squares Deconvolution method to create high signal-to-noise ratio mean Stokes $V$ profiles. We also measured the classical activity indicator S-index for the Ca~{\sc ii} H \& K lines, and the stellar radial velocity (RV).}
{We have unambiguously detected a weak 
Stokes $V$ signal in the spectral lines of Pollux, and measured the related surface-averaged longitudinal magnetic field $B_\ell$.  The longitudinal field averaged over the span of the observations is below one gauss. Our data suggest variations of the longitudinal magnetic field, but no significant variation of the S-index. We observe  variations of RV  which are qualitatively consistent with the published ephemeris for a proposed exoplanet orbiting Pollux. The observed variations of $B_\ell$ appear to mimic those of  RV, but additional data for this relationship to be established. Using evolutionary models including the effects of rotation, we derive the mass of Pollux and we discuss its evolutionary status and the origin of its magnetic field.}
{This work presents the first direct detection of the magnetic field of Pollux, and demonstrates that ESPaDOnS and NARVAL are capable of obtaining sub-G measurements of the surface-averaged longitudinal magnetic field of giant stars, and of directly studying the relationships between  magnetic activity, stellar evolution and planet hosting of these stars.}

   \authorrunning {M. Auri\`ere et al.}
   \titlerunning {Discovery of a weak magnetic field on Pollux}

\maketitle

\section{Introduction}

Pollux ($\beta$ Geminorum, HD 62509, HR 2990, HIP 37826), is a K0III giant neighbour of the Sun (Hipparcos distance of 10.3 pc , $m_V=1.16$~mag). Activity indicators suggest that it may be a weakly-active magnetic star: the core emission of the Ca~{\sc ii} H \& K lines is marginal (at the basal level, Strassmeier et al. 1990) and a long pointed exposure with ROSAT detected weak X-ray emission - 3 orders of magnitude weaker than that of classical magnetically-active giant stars (H\"unsch et al., 1996). Extensive studies of the radial velocity (RV) of Pollux have been performed following the discovery of a RV variation with a total amplitude of about 100 m/s (Walker et al. 1989) and a period of about 580 days (Larson et al. 1993, Hatzes \& Cochran 1993). To interpret these variations, the planetary companion hypothesis is now strongly preferred (Reffert et al. 2006, Hatzes et al. 2006, Han et al. 2008). Furthermore, the discovery of stellar oscillations in Pollux has been recently reported (Hatzes \& Zechmeister 2007). In this context, we have undertaken a very sensitive magnetic study of Pollux using the new-generation twin spectropolarimeters ESPaDOnS at the Canada-France-Hawaii telescope (CFHT) and NARVAL at T\'elescope Bernard Lyot (TBL, Pic du Midi Observatory). We report here observations spanning about 18 months in the 2007/2009 season,  which establish the definite detection of a weak magnetic field at the surface of this star. Section 2 describes our observations, Sect. 3 our results, and Sect. 4 the fundamental parameters of Pollux and its evolutionary status. Section 5 includes a discussion of the nature of the magnetic field of Pollux and Sect. 6 presents our conclusions.

\section{Observations with ESPaDOnS and NARVAL}

Observations of Pollux were obtained at the CFHT using ESPaDOnS (Donati et al., 2006) and at the TBL using NARVAL (Auri\`ere, 2003). ESPaDOnS and NARVAL are twin spectropolarimeters. Each instrument consists of a Cassegrain polarimetric module connected by optical fibres to an echelle spectrometer. In polarimetric mode, the instrument simultaneously acquires two orthogonally-polarized spectra covering the spectral range from 370 nm to 1000 nm in a single exposure, with a resolving power of about 65000. 

A standard circular polarization observation consists of a series of 4 sub-exposures between which the 
 half-wave retarders (Fresnel rhombs) are rotated in order to exchange the paths of the orthogonally-polarized beams within the whole instrument (and therefore the positions of the two spectra on the CCD), thereby reducing spurious polarization signatures. The extraction of the spectra, including wavelength calibration, correction to the heliocentric frame and continuum normalization, was performed using Libre-ESpRIT (Donati et al. 1997), a dedicated and automatic reduction package installed both at CFHT and at TBL. The extracted spectra are output in ascii format, and consist of the normalised Stokes $I$ ($I/I_{\rm c}$) and Stokes $V$ ($V/I_{\rm c}$) parameters as a function of wavelength , along with their associated Stokes $V$ uncertainty $\sigma_V$ (where $I_{\rm c}$ represents the continuum intensity). Also included in the output are "diagnostic null" spectra $N$, which are in principle featureless, and therefore serve to diagnose the presence of spurious contributions to the Stokes $V$ spectrum.

We observed Pollux on 21 different dates between 29 September 2007 and 18 March 2009, acquiring a total of 91 sequences of 4 sub-exposures. Observations at CFHT were part of a "snapshot" program and were sometimes obtained through clouds (data obtained on some of the cloudy nights were discarded because of large and rapid variations in sky transparency induced spurious signatures in the "diagnostic null" spectra $N$). Each spectrum used in this work is of good quality with a peak signal-to-noise ratio (S/N) in Stokes $I$ per 2.6 km s$^{-1}$ spectral bin greater than 1000 (see Table 1). As the rotation period of Pollux is expected to be rather long, and in order to increase the precision of our measurements, the data were time-averaged in 12 groups (each group containing from 4 to 15 Stokes V (and Stokes I) series) which are presented in Table 1. Table 1 comprises the log of observations, and gives the dates, the instrument, the number of averaged Stokes $V$ (and Stokes $I$) series, the mean peak S/N in Stokes I for one reduced spectrum (see above), and the mean Heliocentric Julian Date (HJD) of the binned measurement. Table 1 also reports the phase for the periodic variations of the radial velocity as computed from the ephemeris of Hatzes et al. (2006). In the last row of Table 1 is described the single Stokes V series obtained on 18 October 2008 with ESPaDOnS, that we used for making Fig. 1., for inferring the $^{12}$C/$^{13}$C ratio (Sect. 4.1) and the $v\sin i$ (Sect. 4.2). However, this single ESpaDOnS spectrum has not been averaged with the 30 Sep. 08 NARVAL spectra, and is not used in the variability studies.
To obtain a high-precision diagnosis of the spectral line circular polarization, Least-Squares Deconvolution 
(LSD, Donati et al. 1997) was applied to each reduced Stokes $I$ and $V$ spectrum. We used a solar abundance line mask calculated for an effective 
temperature of 5000 K, $\log g =3.0$, and a microturbulence of 2.0 km s$^{-1}$, consistent with the physical parameters of Pollux (Hekker and Mel\'endez 2007, Takeda et al. 2008). The mean photospheric profile has an effective Land\'e factor of 1.21 and is centred at 564 nm. 

From these mean Stokes profiles we computed the surface-averaged longitudinal magnetic field  $B_\ell$ in G, using the first-order moment method (Rees \& Semel 1979), adapted to LSD profiles (Donati et al. 1997, Wade et al. 2000). These measurements of $B_\ell$ are presented in Table 1 with their 1 $\sigma$ error, in G. These errors are computed from photon statistical error bars propagated through the reduction of the polarization spectra and the computation of the LSD profiles, as described by Wade et al. (2000).

\begin{table*}
\caption{Log of observations of Pollux (for details, see Sect. 2 and Sect. 3 ) }          
\label{table:1}   
\centering                         
\begin{tabular}{c c c c c c c c c c}     
\hline\hline               
Date          &Instrument& Stokes V  & S/N     &  HJD      &Phase & $B_\ell$ & $\sigma$ & $S$-index     &RV \\
              &          &           &         &(245 0000+)&       & G     & G      & Ca~{\sc ii} & km s$^{-1}$ \\
\hline                        
29 Sep - 02 Oct 07 & ESPaDOnS &4          & 1196    & 4375.1    & 0.25  &-0.67   & 0.26    & 0.1090       & 3.651 \\
12 Dec - 18 Dec 07 & NARVAL   &6          & 1792    & 4450.7    & 0.38  &-0.51   & 0.14    & 0.1234       & (3.557)   \\
31 Dec 07     & ESPaDOnS &5          & 1327    & 4467.0    & 0.40  &-0.57   & 0.19    & 0.1081       & 3.538    \\
05 Apr 08     & NARVAL   &4          & 1986    & 4562.4    & 0.57  &-0.15   & 0.15    & 0.1263       & 3.532    \\
15 Apr 08     & NARVAL   &8          &  960    & 4572.4    & 0.58  &-0.12   & 0.23    & 0.1267       & 3.508    \\
16 Sep 08     & NARVAL   &8          & 1566    & 4726.7    & 0.85  &-0.45   & 0.13    & 0.1249       & 3.551    \\
20 Sep - 21 Sep 08 & NARVAL   &8          & 1328    & 4731.2    & 0.86  &-0.21   & 0.16    & 0.1262       & 3.576    \\
30 Sep 08     & NARVAL   &8          & 1955    & 4740.7    & 0.87  &-0.27   & 0.11    & 0.1240       & 3.584    \\
21 Dec 08     & NARVAL   &8          & 1532    & 4822.6    & 0.01  &-0.33   & 0.14    & 0.1255       & 3.616    \\
25 Feb 09     & NARVAL   &8          & 1934    & 4888.3    & 0.12  &-0.59   & 0.11    & 0.1259       & 3.633 \\
12 Mar 09     & NARVAL   &8          & 1190    & 4903.4    & 0.15  &-0.69   & 0.14    & 0.1239       & 3.640 \\
18 Mar 09     & NARVAL   &15         & 1528    & 4909.4    & 0.16  &-0.56   & 0.10    & 0.1225       & (3.634) \\
18 Oct 08     & ESPaDOnS &1          & 1839    & 4759.1    & 0.90  &-1.40   & 0.35    & 0.1102       & 3.586    \\
\hline
\hline                            
\end{tabular}
Notes:Individual columns report dates of observation, instrument, number of averaged Stokes $V$ series, mean peak S/N in Stokes $I$ for one spectrum, mean HJD, Phase according to Hatzes et al. 2006, $B_\ell$ and its error in G, the $S$-index for Ca~{\sc ii} H and K, and the radial velocity measured from the LSD Stokes $I$ profile.  
\end{table*} 

\section{Results of the observations}

\subsection{Direct detection of a weak magnetic field on Pollux}

A characteristic magnetic Stokes $V$ Zeeman signature appears in nearly all the individual Stokes $V$ LSD spectra. As ESPaDOnS and NARVAL are twin instruments and as no large variation of the Stokes $V$ signature is observed (see Sect. 3.2), we averaged the 91 LSD profiles to produce global mean Stokes $I$, $V$ and $N$ profiles to obtain the highest S/N ratio possible, and applied to them the LSD statistical detection criteria (Donati et al., 1997).
\begin{figure}
\centering
\includegraphics[width=9 cm,angle=0] {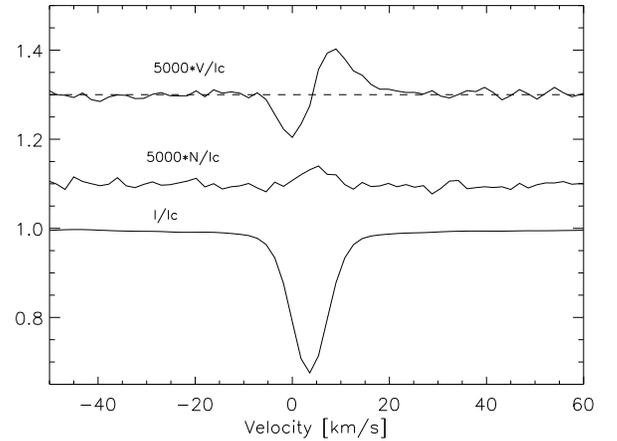} 

\caption{Mean LSD profiles of Pollux from all 91 spectra acquired with ESPaDOnS and NARVAL in the September 2007 - March 2009 period. From top to bottom are Stokes $V$, null polarisation $N$, and Stokes $I$ profiles. For display purposes , the profiles are shifted vertically, and the Stokes $V$ and diagnostic $N$ profiles are expanded by a factor of 5000. The dashed line illustrates the zero level for the Stokes $V$ profile.}
\end{figure}

 The global mean LSD profiles obtained are presented in Fig. 1. The average Stokes $V$ profile shows a definite Zeeman detection with a reduced $\chi^{2}$ equal to 5.79 . The polarisation signal is weak with an amplitude of  $2\times 10^{-5}$ of the continuum. The corresponding surface-averaged longitudinal magnetic field is $B_\ell$ = -0.46 $\pm$ 0.04 G. The associated error bar is the formal root mean square error associated with the global mean LSD profiles.

\subsection{Variations of radial velocity, magnetic field and activity indicators}

Clear variations of radial velocity of Pollux reported in the literature are interpreted as the result of the presence of an exoplanet (Reffert et al. 2006, Hatzes et al. 2006, Han et al. 2008). Our magnetic study is critical in this context, as magnetic activity represents a credible alternative to the exoplanet hypothesis (Queloz et al., 2001). 

 The radial velocity RV of Pollux was measured from the averaged LSD Stokes $I$ profiles using a Gaussian fit. The radial velocity stability of ESPaDOnS and NARVAL is about 20-30 ms$^{-1}$ (Moutou et al. 2007). Table 1 presents our measurements of RV as well as the phase computed from the ephemeris of Hatzes et al. (2006; phase 0 corresponds to the maximum RV in their observations (periastron); HJD$_0$=2447739.02, $P=589.64$~days). The RV reported for 12-18 December 2007 is that obtained for 18 December and is given between brackets because the measurements for the other dates are significantly (about 100 ms$^{-1}$) smaller than the typical value for Pollux. The same is true for 18 March 2009, when only 9 about the 15 RV measurements could be used.
 In the lower frame of Fig. 2 we show the variations of RV with respect to HJD (constant error bars of 30 ms$^{-1}$ are shown). Table 1 shows that the observed variations are qualitatively consistent with the ephemeris of Hatzes et al. (2006): higher values occur near phase 0, while smaller values occur near phase 0.5, and the total amplitude is near 100 ms$^{-1}$. However, the best fit of our RV data with the Hatzes et al. period results in an 80-day shift of HJD$_0$ with respect to that reported by Hatzes et al.

$B_\ell$ remains of negative polarity for all our observations. Table 1 suggests that the unsigned value of the longitudinal magnetic field decreases towards a minimum in April 2008, then increases. Averaging all the LSD Stokes $V$ profiles of 2007 and all those of April 2008 we obtain corresponding values of $B_\ell$ of respectively -0.56 $\pm$0.1 G (the same value as for 18 March 2009) and -0.12$\pm$ 0.13 G. The error bars are the root mean square errors associated with averaging the data in the group. The variation of $B_\ell$ between the 2007 and 18 March 2009 observations, and the April 2008 observations, is significant at the 2.75 $\sigma$ level. Figure 3 illustrates the variations of the averaged Stokes $V$ profiles for the 12 groups of observations reported in Table 1.
 
The upper frame of Fig. 2 shows the variation of $B_\ell$ with respect to HJD (1$\sigma$ error bars are shown); these variations of $B_\ell$ appear to mimic the RV variations. Figure 4 plots the variation of $B_\ell$ and that of RV. The correlation factor between $B_\ell$ and RV measurements inferred from a linear regression is -0.67, and is therefore not insignificant. This correlation is discussed in the next subsection.

In order to monitor the line-activity indicators, we computed the $S$-index (defined from the Mount Wilson survey, Duncan et al. 1991) for the chromospheric Ca~{\sc ii} H \& K line cores. Our procedure was first calibrated on the main sequence solar-type stars of Wright et al. (2004), then we added 0.03 to the index to fit measurements of 5 giant stars observed by Duncan et al. (1991) and Young et al. (1989). Our averaged values of the $S$-index of Pollux are presented in Table 1. They are rather weak and consistent with the 2 measurements presented by Duncan et al. (1991). They are also consistent with the McDonald $S$-index values of Pollux presented by Hatzes et al. (2006). We do not observe significant variations of the $S$-index, apart from a systematic shift between NARVAL and ESPaDOnS values (about 0.016), which could be due to small continuum normalisation differences. The absence of a significant $S$-index variation is consistent with the weak activity level of Pollux.

\subsection{A possible correlation between $B_\ell$ and RV }

 One conclusion from the previous section is that there may exist a correlation between the $B_\ell$ and RV measurements. Hatzes (1999) devised a simple model to infer the effect of magnetic spots on RV variations. His study suggests that the RV variations due to spots are comparable in magnitude to $v\sin i$. Therefore, as the $v\sin i$ of Pollux is low, a magnetic spot could only explain a small part of the observed RV variation. However, these simulations are from simple modelling and need to be investigated further (Hatzes, 1999). For the active red giant EK Eri, a clear correlation between $B_\ell$ and RV is observed (Auri\`ere et al. 2008) which  supports the suggestion that the RV variations are due to magnetic activity (Dall et al. 2005). In the case of the main sequence star HD 166435, it has been concluded that the observed periodic variations of RV (about 200 m/s total amplitude) could be completely explained by magnetic spots (Queloz et al. 2001, Martinez Fiorenzano et al. 2005). Now, relations between hot Jupiters and stellar activity have been observed in several systems (Lanza, 2008); in these cases the hosting stars are on the main sequence and the orbital periods are short. All these results show that it will be important to observe Pollux with ESPaDOnS and NARVAL during more than one period of RV variation to settle the correlation with $B_\ell$. Then, it will be possible to discuss more deeply the relation between magnetic activity and RV variations of this star.

\begin{figure}
\centering

\includegraphics[width=6 cm,angle=0] {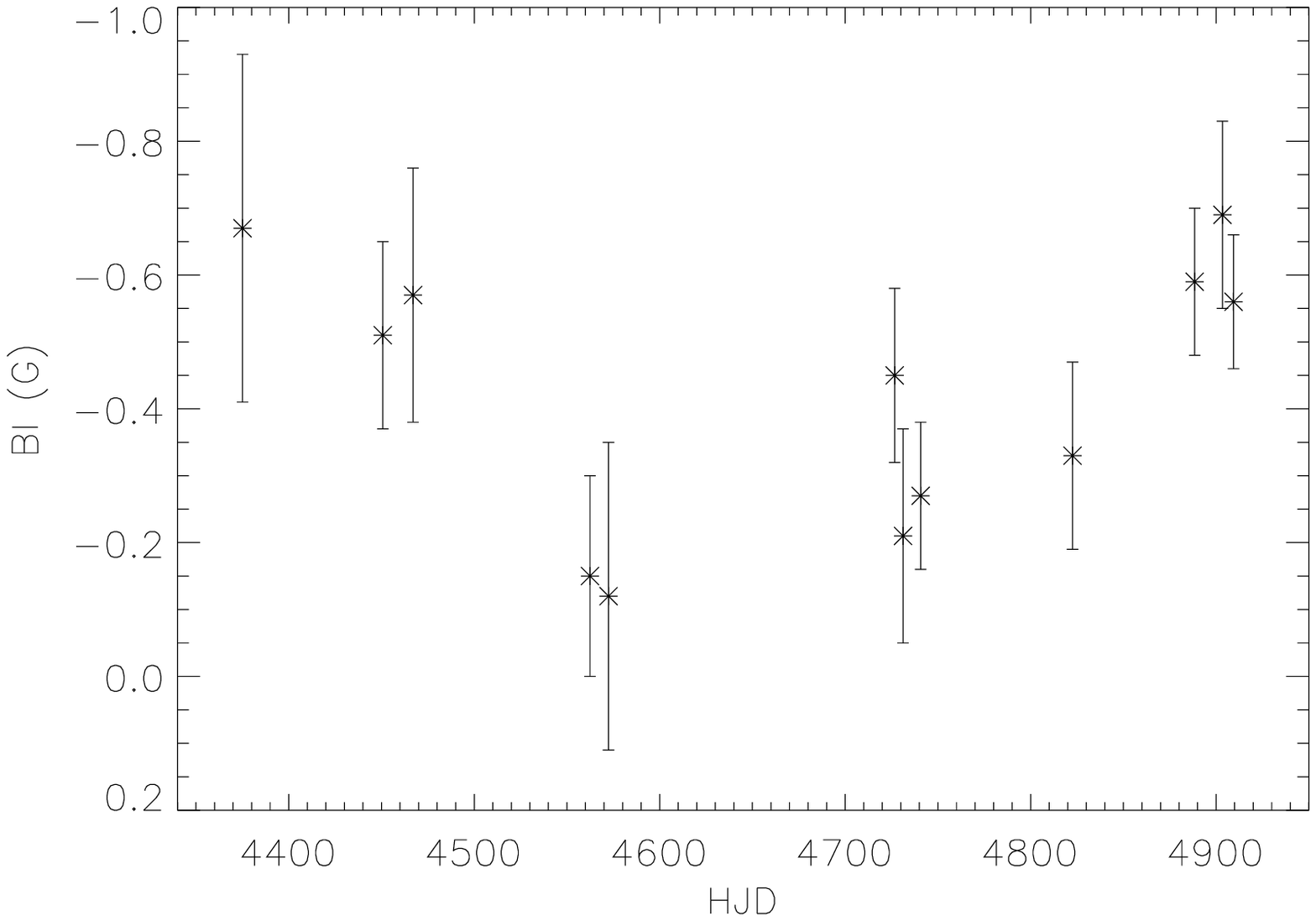}
\includegraphics[width=6 cm,angle=0] {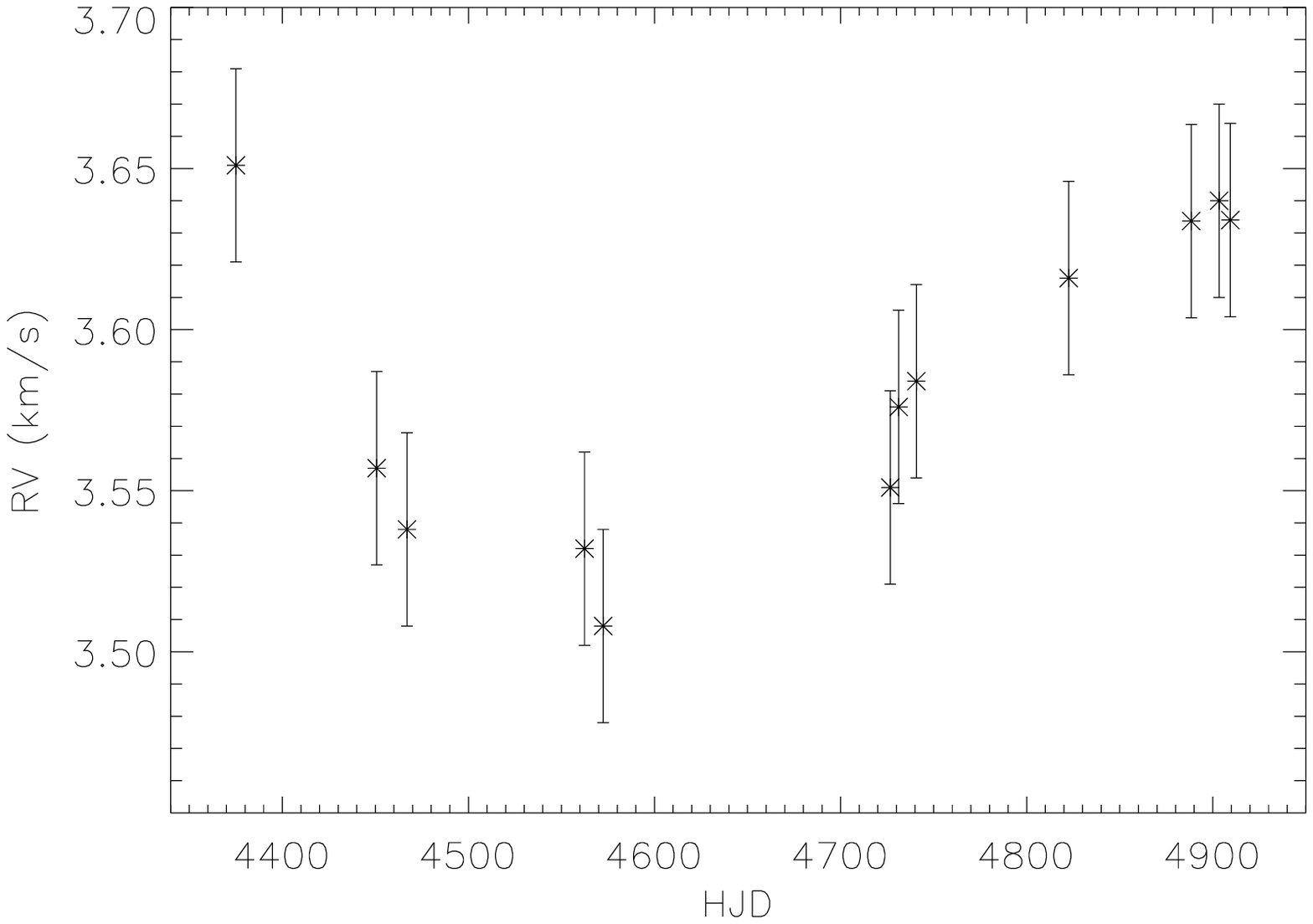}

\caption{Variations of radial velocity and longitudinal magnetic field of Pollux with HJD. X-axis: HJD - 2450000. Bottom: RV in km s$^{-1}$ (error bars: 0.030 km s$^{-1}$); top: $B_\ell$ in G (ticks show 1 $\sigma$ error bars).}
\end{figure}

\begin{figure}
\centering
\includegraphics[width=8 cm,angle=0] {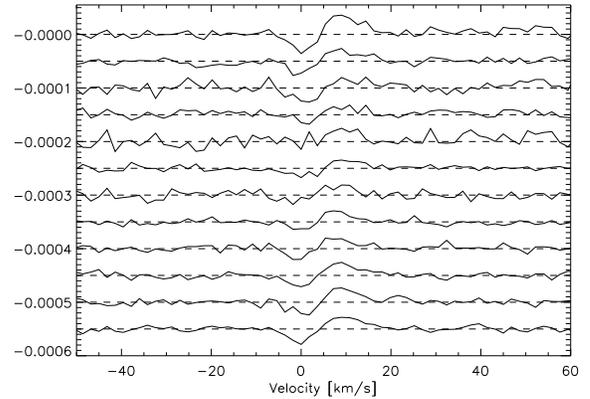}

\caption{Variation of averaged LSD Stokes $V$ profiles of Pollux. From top to bottom (as in Table 1): 29 Sep. - 02 Oct. 07, 12 Dec. - 18 Dec. 07, 31 Dec. 07, 05 Apr. 08, 15 Apr. 08, 16 Sept. 08, 20 Sep. - 21 Sep. 08, 30 Sep. 08, 21 Dec. 08, 25 Feb. 09, 12 Mar. 09, 18 Mar. 09. For display purposes, each profile is shifted by 0.00005.  The dashed lines illustrate the zero level for the Stokes $V$ profiles.}
\end{figure}

\begin{figure}
\centering
\includegraphics[width=8 cm,angle=0] {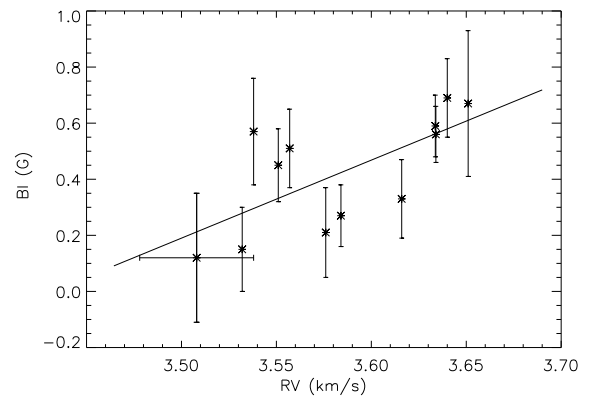}

\caption{Variations of longitudinal magnetic field $B_l$ of Pollux with respect to radial velocity RV. The straight line corresponds to a linear regression. $B_\ell$ is in units of G (ticks show 1 $\sigma$ error bars), while RV is in units of km s$^{-1}$. The horizontal ticks of 0.03 km/s shown for the first point on the left illustrate the error in RV. For details see Sect. 3.2.}
\end{figure}

\section{Fundamental parameters and evolutionary status of Pollux}

As Pollux is a bright, nearby, and well-studied star with an Hipparcos distance and an 
interferometric radius measurement (Nordgren et al., 2001), it should be a good case 
for precise determination of its fundamental parameters. 
However, published values of $T_{\rm eff}$ range from 4925-4660 K, log$g$ from 3.15-2.52, 
metallicity [Fe/H] from -0.07 to +0.19, and mass from 1.7-2.3~$M_{\odot}$. 
In Table 2 we present some parameters for Pollux from the more commonly cited papers in the modern literature. 
References are those given at the end of the paper.

\begin{table}
\caption{Physical parameters of Pollux from the literature}           
\label{table:2}   
\centering                         
\begin{tabular}{c c c c c l}     
\hline\hline               
$T_{\rm eff}$  & $\log g$ & Fe/H   & vsini      &Mass        & References   \\
           &          &        & km s$^{-1}$& $M_{\odot}$&             \\
\hline 
           &          &        & 2.5        &            & Gray 1982    \\
4865       &2.75      &        &            &1.7         & Drake 1991   \\
           &          &        & 1.7        &            & Fekel 1997    \\
4850       & 2.96     & -0.07  &            &            & McWilliams 1990  \\
4786       & 2.77     &        &            &1.7         & Allende-Prieto 1999   \\
4850       & 2.52     & +0.08  &            &            & Gray 2003      \\
4660       & 2.68     & 0.19   &            &            & Allende-Prieto 2004   \\
4925       & 3.15     &        & 1.67       &            & Hekker 2007    \\
4904       & 2.84     & +0.04  & 1.61       &2.31        & Takeda 2008    \\
4841       &          &        &  2.7       &            & Massarotti 2008 \\

\hline                                 
\end{tabular}
\end{table} 

\subsection {The mass of Pollux and its evolutionary status}

Knowledge of the mass of Pollux is important both for understanding its evolutionary 
status and for inferring the mass of its hypothetical planetary companion. 
Figure 5 shows the position of Pollux in the HR diagram, using the luminosity deduced from the Hipparcos parallax 
and  $T_{\rm eff}$ determinations by different authors.
Among the most recent published values, we note that the effective temperatures derived by Takeda et al. (2008) 
or Massarotti et al. (2008) are in better agreement with the $T_{\rm eff}$ 
one obtains when assuming the Hipparcos luminosity and the radius of 8.8 $\pm$ 0.1 $R_{\odot}$ determined from 
the interferometric angular diameter of Nordgren et al. (2001)
evaluated at the Hipparcos distance. However we also show the low $T_{\rm eff}$ value derived by Allende-Prieto et al. (2004) for illustration of 
the uncertainties.

Figure 5 shows the position of Pollux in the HR diagram, where it is compared with standard evolutionary tracks (ignoring stellar rotation), as well as those computed with the STAREVOL code (including stellar rotation),
all assuming solar metallicity (with the composition of Asplund et al. 2005) and various initial masses (Charbonnel et al., in preparation).  This comparison indicates an initial mass of Pollux of either $2.55 \pm 0.25$ $M_{\odot}$ or $2.4 \pm 0.25$ $M_{\odot}$, 
when the Takeda or Massarotti $T_{\rm eff}$s are considered respectively.
According to this diagram, the star appears to be either at the base of the red giant branch (RGB), or at the clump (i.e. burning He in its core) 
as already suggested by Drake \& Smith (1991).
Our theoretical mass is consistent with the mass range of 2.04 $\pm$ 0.3 M$_{\odot}$ inferred by Hatzes and Zechmeister (2007) from their asteroseismic study. 

Let us note that the effective temperature determined by Allende-Prieto et al. (2004) implies masses of 2.0 $\pm$ 0.2 M$_{\odot}$
when compared to our tracks. However, in this case the theoretical stellar radius is much larger (i.e., 9.49 R$_{\odot}$) compared to the Nordgren interferometric measurement.

For a 2.5~M$_{\odot}$ star, three models including the effects of stellar rotation (i.e., the transport of angular momentum and of chemicals 
by meridional circulation and shear-induced turbulence, see e.g. Decressin et al. 2009) are also shown. 
These models assume initial rotational velocities of 50, 100, and 180  km\,s$^{-1}$ that are characteristic of the possible A-type main sequence progenitors of Pollux (see sect. 5.2). 
Figure 5 shows that our determination of the mass and of the evolutionary status 
of Pollux does not change when rotation is included. 

More information about Pollux
can be inferred from its surface Li abundance. 
Values of N(Li)\footnote{N(Li) = log$_{10}$ (A(Li)) + 12} equal to 0.6 and 0.83 
are reported by Brown et al. (1989) and Mallik (1999) respectively. 
Fig. 6 shows how these measurements fit with predictions of our models for a star of 2.5~$M_{\odot}$ and different initial rotations.
It suggests that the initial (i.e., ZAMS) rotation velocity of Pollux was of the order of $75\pm 25$~km\,s$^{-1}$. The Li abundance does not allow us however to further constrain the evolutionary status of Pollux, since this quantity is not expected to change between the end of the first dredge-up phase (around $T_{\rm eff}$ $\sim$ 4700~K) and the clump in a 2.5~$M_{\odot}$ star (e.g., Charbonnel \& Zahn 2007).

A precise determination of the carbon isotopic ratio of Pollux appears to be more promising in this respect. Indeed, at  $T_{\rm eff}$ = 4900~K on the early-RGB, the first dredge-up is not completely 
finished, and the theoretical $^{12}$C/$^{13}$C ratio predicted by our 
2.5~$M_{\odot}$ models is $\sim$ 28 . At the same $T_{\rm eff}$ but on 
the clump (i.e., after the completion of the first dredge-up), the 
$^{12}$C/$^{13}$C ratio is expected to be $\sim$ 20. These values are 
from the standard model. Including a rotation of 75 km/s gives similar values for the $^{12}$C/$^{13}$C ratio, i.e. 27 on the early-RGB and 18 on the clump . 
We have therefore measured our spectrum obtained on 18 October 2008 to investigate this isotopic ratio. 

Abundances were derived using MARCS model atmospheres  (Gustafsson et al. 2008) and the LTE spectral synthesis code TurboSpectrum (Alvarez \& Plez 1998). The $^{12}$C/$^{13}$C ratio was determined on the basis of $^{12}$CN and $^{13}$CN features in the $\lambda$ 800-800.6 nm  wavelength range. 
The CN molecular data (line positions, excitation values, and oscillator strengths) are taken from the investigation of de Laverny \& Gustafsson (1998; private communication from P. de Laverny 2009). 
Using a MARCS model atmosphere with $T_{\rm eff}$ = 5000 K, $\log g$ = 3.0, mass = 2 $M_{\odot}$, we derived $^{12}$C/$^{13}$C ratio = 24. Including all the uncertainties (temperature, gravity, 
 microturbulence, continuum determination) we consider that the lower and upper limits on the 
$^{12}$C/$^{13}$C ratio are 19 and 39, respectively. 
This is unfortunately insufficiently precise to enable us to derive the $^{12}$C/$^{13}$C ratio with an accuracy able to discriminate between the RGB and clump hypothesis. 

In the RGB and clump hypothesis, the theoretical depths of the convective zone of the 2.5~$M_{\odot}$ star 
are $M_{\rm r}= 1.58$ and 1.55~$M_{\odot}$ respectively. For the same evolutionary model, the  turnover time is 170 days (respectively 176 days) in the RGB (respectively clump) hypothesis, at $H_P/2$ above the base of the convective zone (where $H_P$ is the pressure scale height). The data we have used in this paper are consistent with both Pollux being at the base of the RGB or at the clump. This ambiguity will not hamper our discussion in the following sections.

\begin{figure}
 \centering
\includegraphics[width=8 cm,angle=0] {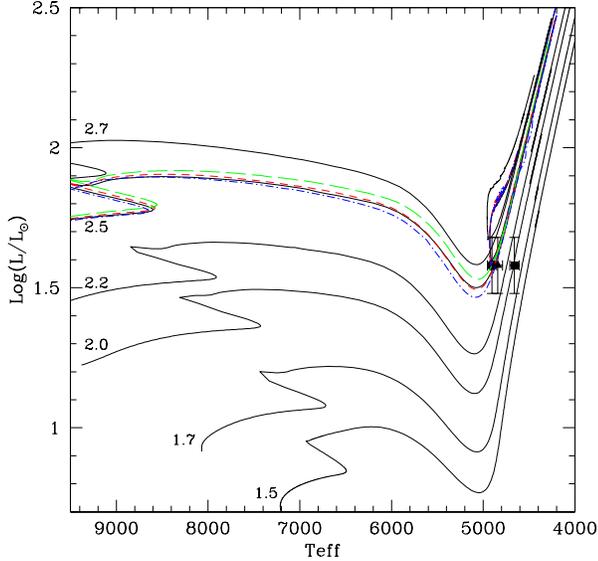} 

\caption{HR diagram showing the position of Pollux for observational values of $T_{\rm eff}$
obtained from Takeda et al.~(2008; black point, on the left), Massarotti et al. (2008; black triangle, middle), 
and Allende-Prieto et al. (2004; black square, on the right), and for the Hipparcos luminosity.
The observational uncertainties are 50~K for $T_{\rm eff}$ (approximately the diameter of the symbol) and 0.1 for log ($L/L_{\odot}$) (tick). 
Standard evolutionary tracks for solar composition and for various initial stellar masses as indicated are plotted (solid lines). 
For 2.5~M$_{\odot}$, three models with rotation are also shown, corresponding to initial rotational velocities of 50, 100, and 180 km\,s$^{-1}$ 
(coloured lines, from bottom to top respectively). The He-burning phase and the early-AGB are drawn only for the 2.5 and 2.7~M$_{\odot}$ models. For details see sect. 4.1.}

\label{f3}
\end{figure}

\begin{figure}
 \centering
\includegraphics[width=8 cm,angle=0] {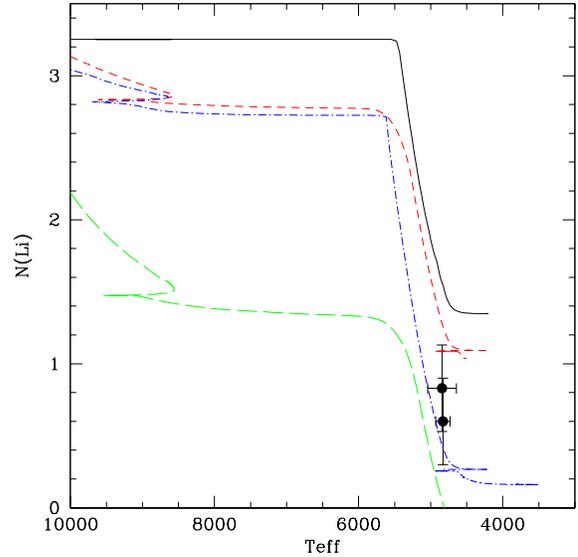} 

\caption{Lithium abundance versus effective temperature measured in Pollux by Brown et al. (1989) and Mallik (1999) 
(Here we show the $T_{\rm eff}$ used in the abundance analyses, i.e., 4830 and 4843 K respectively). 
The tracks show the theoretical evolution of the surface Li abundance up to the AGB phase for a star of mass 2.5~$M_{\odot}$ 
of solar composition in the cases without rotation (continuous black line) and  with initial rotational velocities of 50, 100, 
and 180 km\,s$^{-1}$ (red, blue, and green lines, from top to bottom, respectively).  
First dredge-up ends around $T_{\rm eff}$ $\sim$ 4700~K.
For details see Sect. 4.1.}
\label{f3}
\end{figure}

\subsection {The rotational velocity and period of Pollux}

Knowing the rotational period of Pollux is critical for inferring the origin of the surface magnetic field. In addition to the 589.84 day period of the RV variation, a period of about 130 days is derived from Hipparcos photometry (Hatzes et al. 2006) as a possible rotational period. As the radius of Pollux is known (8.8 $\pm$ 0.1 $R_{\odot}$; Nordgren et al., 2001), the projected rotational velocity $v\sin i$ and the rigid rotation hypothesis can be used to derive an upper limit on the rotational period. For example, establishing that $v\sin i > 0.75 $ km\,s$^{-1}$would exclude the 589.64 day period (and longer periods). Using the CORAVEL radial velocity scanner, De Medeiros \& Mayor (1999) inferred $v\sin i<1$~km\, s$^{-1}$. However, the $v\sin i$ values reported in Table 2 are all greater than 1~km\, s$^{-1}$. 

Because ESPaDOnS and NARVAL have a spectral resolution (about 65000) equivalent to
the best used in the recent studies cited in Table 2, we have performed our own evaluation of the $v\sin i$
using the spectrum observed with ESPaDOnS on 18 October 2008. We used a direct
fitting method assuming a gaussian instrumental broadening function, with a FWHM
of 4.4 km s$^{-1}$. The {\sc Zeeman2} code (Landstreet 1988, Wade et al. 2001) was used to compute synthetic spectra of Pollux, taking into account $v\sin i$ and
macroturbulence, for which a radial-tangential parametrisation was used. An ATLAS9 model atmosphere with solar metallicity, an effective temperature of 4904~K and a logarithmic surface gravity of 2.84
was employed. The synthetic spectrum was then fitted to the observed spectrum in the 550-560 nm and 600-610 nm ranges, using a Levenberg-Marquardt $\chi^2$ minimisation procedure (Press et al. 1992).
Our main result is that there is a degeneracy between macroturbulence and $v\sin i$, i.e. macroturbulence of 4 km s$^{-1}$ and $v\sin i = 0$ km\,s$^{-1}$ (best fit) and macroturbulence of 0 km s$^{-1}$ and $v\sin i$ = 5.3 km\,s$^{-1}$ are both  possible solutions. It is known that macroturbulence is seldom absent in red giants, and can reach values higher than 5 km\,s$^{-1}$ (Carney et al. 2008). It is generally found to be in the range 3-4 km\,s$^{-1}$ for Pollux (Gray 1982, Carney et al. 2008, Hekker and Melendez 2007). Macroturbulence therefore likely dominates the measured broadening observed for Pollux and our resolution of 4.4 km s$^{-1}$ appears insufficient to disentangle the $v\sin i$ contribution. In conclusion, we consider that with the spectral resolution used in the recent studies of Pollux, small values of $v\sin i$ cannot be excluded, nor can rotational periods as large as the 589.64 d period observed in the RV variations.

\section {The magnetic field of Pollux}

\subsection {Magnetic field and activity of Pollux}

  Pollux presents weak chromospheric emission at a basal level: this could be due to accoustic or hydrodynamic heating rather than to magnetic activity (Hall 2008). The direct detection of a surface magnetic field now points to magnetism as the probable driver. 
Figure 1 and Fig. 3 show the Stokes $V$ feature typical of a simple magnetic topology or a large-scale field. {However, because of the small $v\sin i$ of Pollux, the Stokes $V$ profile does not resolve complex structures which might be associated with dynamo action.
 
Now we consider the question of the nature and origin of this weak magnetic field. Strong magnetic fields have been observed at the surfaces of rapidly-rotating giants, including RS CVn binaries (Berdyugina et al., 2006) or FK Com stars (e.g. HD 199178, Petit et al. 2004). Weaker magnetic fields ($B_\ell$ of the order of a few G) have been detected with NARVAL or ESPaDOnS at the surface of active single giants, which rotate faster than the bulk of the red giant class (Konstantinova-Antova et al. 2008, 2009). For all these active stars, the origin of the magnetic field is very likely a dynamo. The exceptional, strong, magnetic field at the surface of the slowly-rotating active giant EK Eri has been recently detected with NARVAL (Auri\`ere et al., 2008). That magnetic study supports the field being the descendant of that of a strongly magnetic Ap star (Stepie\'n, 1993, Strassmeier et al. 1999). 

The magnetic field we observe at the surface of Pollux is one of the weakest ever measured on a star: it is at the level of, or even weaker than, that which is observed at the surface of solar twins (Petit et al., 2008). This confirms the activity level observed with ROSAT (Schr\"oder et al., 1998), an X-ray luminosity of about $10^{27}$ erg\,s$^{-1}$, i.e. the average level of the X-ray luminosity of the Sun (Katsova \& Livshits, 2006).

\subsection {Origin of the magnetic field}

The main sequence progenitor of Pollux was approximately a 2.5 M$_{\odot}$, A2V star (Sect. 4.1 and Allen, 2000). Its ZAMS radius would have been about 2 $R_{\odot}$. Such a progenitor could have been either a 'normal' A-type stars (which are rather fast rotators), or a magnetic Ap star (which represent about 5$\%$ of main sequence A-type stars and which are slow rotators, with equatorial velocity generally smaller than 100 km s$^{-1}$, Abt \& Morrell, 1995). 
From Sect. 4.1 , fitting the Li abundance with our evolutionary models, we infer that the progenitor of Pollux was a moderate rotator, which is consistent both with 'normal' A-type star and magnetic Ap origins.  

\subsubsection {Pollux as the descendant of a 'normal' A-type star}

Normal A-type stars have weak surface convection on the main sequence and appear to be free of large scale surface magnetic fields (Shorlin et al. 2002 - but see also Ligni\`eres et al. 2009). During the subgiant phase, a convection zone appears and deepens and, with the combination of the high rotation rate, a dynamo magnetic field is expected to occur.
 At the evolutionary phase of Pollux on the RGB, our evolutionary model predicts 
a convective turnover time of about 170 days at $H_P/2$ above the bottom of the convective zone (see Sect. 4.1). As the rotation period is expected to be rather smaller than some hundreds of days (see Sect. 4.2), the Rossby number would be of the order of 1 for Pollux, and the requirements for a dynamo might be fulfilled (Durney and Latour 1978).

\subsubsection {Pollux as the descendant of an Ap star}

The mass of Pollux corresponds to about the peak of the mass incidence distribution of magnetic Ap stars in the solar neighbourhood (Power et al., 2007 and in prep.). Magnetic Ap stars are slow rotators with respect to 'normal' A-type stars (Abt \& Morrell, 1995) and harbour organised surface magnetic fields which can be roughly modeled by a dipole (Landstreet, 1992). Following the calculations of Stepie\'n (1993), assuming only conservation of magnetic flux during stellar evolution, we find that the surface magnetic field of a magnetic Ap star (a main sequence star progenitor of radius 2 $R_{\odot}$, with a magnetic field typical of the nearby Ap stars ($\sim 2.5$~kG)), the surface field at the evolutionary phase of Pollux would be about 130~G - much stronger than that measured. On the other hand, if the progenitor's magnetic field was much weaker - near the field threshold of about 300~G found by Auri\`ere et al. (2007) - the field would be just 15 G at the evolutionary phase of Pollux. This would give a peak $B_\ell$ smaller than 5 G, and would be roughly compatible with our Pollux measurements.

\section{Conclusion}

The weak magnetic field detected and measured at the surface of Pollux may be generated by a dynamo occuring naturally during the evolution of a 'normal' 2.5 M$_{\odot}$ star. Alternatively, it could be the result of the evolution of the large scale fossil magnetic field of an Ap star. However, such a fossil field would necessarily be very weak, near the 300~G field threshold reported by Auri\`ere et al. (2007). As Ap stars hosting such weak fields appear to be quite rare (Auri\`ere et al. 2007, Power et al. 2007 and in prep.), the latter scenario appears less likely than the former.
Whatever the field origin, follow-up observations are required to confirm the  measured variations of the magnetic field and to infer the rotation period of Pollux. If correlated with the radial velocity variations, variations of the magnetic field would enable to discuss the relation between RV variations, magnetic activity and the hypothetical exoplanet.

The sensitivity of ESPaDOnS and NARVAL enables us to perform magnetic studies measuring surface-averaged longitudinal fields smaller than 1 G on red giants, i.e. to study magnetic activity at a level similar to that encountered in the Sun. We are thus in a position to complement X-ray surveys (Schr\"oder et al., 1998) in studying stellar magnetic activity in advanced evolutionary stages of the red giant branch. Magnetic fields are also measured on the AGB, in SiO and OH masers (Herpin et al. 2006; Fish et al. 2006). Using NARVAL, a magnetic field of a few G has been recently detected at the surface of the rapidly-rotating single M5 AGB star EK Boo (Konstantinova-Antova et al., 2009). AGB stars in general may be very slow rotators and other dynamo regimes than that which occurs in the Sun may be invoked there. Along the red giant branch, the dominant dynamo mode may evolve from a boundary-layer dynamo mode to a deep-envelope dynamo mode, and then to a turbulent dynamo mode, when the convective zone deepens and rotation slows (Schrijver and Zwaan, 2000).

\begin{acknowledgements}
       We thank  the TBL and CFHT teams for providing service observing, and Nicole Letourneur, Jean Pierre Michel and Jean-Sebastien Devaux for making observations with NARVAL. The observations obtained with NARVAL  in 2008 were supported by the OPTICON trans-national access program. We thank P. de Laverny for providing us with his CN lines list. WW acknowledges support from the Austrian Science Fond (P17890). GAW acknowledges support from the Natural Science and Engineering Research Council of Canada (NSERC). R. K.-A. acknowledges partial support by the Bulgarian NSF grant DO 02-85 (CVP01/002).
\end{acknowledgements}

\end{document}